\newcommand{\BP}{$G_{\rm BP}$ }
\newcommand{\RP}{$G_{\rm RP}$ }
\newcommand{\Gaia}{\emph{Gaia}}
\newcommand{\Hip}{\emph{Hipparcos}}
\newcommand{\GDR}{\Gaia~DR}
\newcommand\tyctwo{Tycho-2}
\def\gmag{$G$}

\documentclass{aa}  
\usepackage{graphicx}
\usepackage{txfonts}
\usepackage{hyperref}
\begin{document} 
   \title{Gaia data release 1}
   \subtitle{The Photometric Data}	
\author{
 F.        ~van Leeuwen                   \inst{\ref{inst:0001}}
\and D.W.      ~Evans                         \inst{\ref{inst:0001}}
\and F.        ~De Angeli                     \inst{\ref{inst:0001}}
\and C.        ~Jordi                         \inst{\ref{inst:0014}}
\and G.        ~Busso                         \inst{\ref{inst:0001}}
\and C.        ~Cacciari                      \inst{\ref{inst:0043}}
\and M.        ~Riello                        \inst{\ref{inst:0001}}
\and E.        ~Pancino                       \inst{\ref{inst:0024},\ref{inst:0122}}
\and G.        ~Altavilla                     \inst{\ref{inst:0043}}
\and A.G.A.    ~Brown                         \inst{\ref{inst:0011}}
\and P.        ~Burgess                       \inst{\ref{inst:0001}}
\and J.M.      ~Carrasco                      \inst{\ref{inst:0014}}
\and G.        ~Cocozza                       \inst{\ref{inst:0043}}
\and S.        ~Cowell                        \inst{\ref{inst:0001}}
\and M.        ~Davidson                      \inst{\ref{inst:0091}}
\and F.        ~De Luise                      \inst{\ref{inst:0222}}
\and C.        ~Fabricius                     \inst{\ref{inst:0014}}
\and S.        ~Galleti                       \inst{\ref{inst:0043}}
\and G.        ~Gilmore                       \inst{\ref{inst:0001}}
\and G.        ~Giuffrida                     \inst{\ref{inst:0122}}
\and N.C.      ~Hambly                        \inst{\ref{inst:0091}}
\and D.L.      ~Harrison                      \inst{\ref{inst:0001},\ref{inst:0104}}
\and S.T.      ~Hodgkin                       \inst{\ref{inst:0001}}
\and G.        ~Holland                       \inst{\ref{inst:0001}}
\and I.        ~MacDonald                     \inst{\ref{inst:0091}}
\and S.        ~Marinoni                      \inst{\ref{inst:0201},\ref{inst:0122}}
\and P.        ~Montegriffo                   \inst{\ref{inst:0043}}
\and P.        ~Osborne                       \inst{\ref{inst:0001}}
\and S.        ~Ragaini                       \inst{\ref{inst:0043}}
\and P.J.      ~Richards                      \inst{\ref{inst:0059}}
\and N.        ~Rowell                        \inst{\ref{inst:0091}}
\and H.        ~Voss                          \inst{\ref{inst:0014}}
\and N.A.      ~Walton                        \inst{\ref{inst:0001}}
\and M.        ~Weiler                        \inst{\ref{inst:0014}}
\and M.        ~Castellani                    \inst{\ref{inst:0201}}
\and A.        ~Delgado                       \inst{\ref{inst:0001}}
\and E.        ~H{\o}g                        \inst{\ref{inst:0035}}
\and M.        ~van Leeuwen                   \inst{\ref{inst:0001}}
\and N.R.      ~Millar                        \inst{\ref{inst:0001}}
\and C.        ~Pagani                        \inst{\ref{inst:0165}}
\and A.M.      ~Piersimoni                    \inst{\ref{inst:0222}}
\and L.        ~Pulone                        \inst{\ref{inst:0201}}
\and G.        ~Rixon                         \inst{\ref{inst:0001}}
\and F.F.      ~Suess                         \inst{\ref{inst:0001}}
\and \L{}.     ~Wyrzykowski                   \inst{\ref{inst:0001},\ref{inst:0498}}
\and A.        ~Yoldas                        \inst{\ref{inst:0001}}
\and A.        ~Alecu                         \inst{\ref{inst:0001}}
\and P.M.      ~Allan                         \inst{\ref{inst:0059}}
\and L.        ~Balaguer-N\'{u}\~{n}ez        \inst{\ref{inst:0014}}
\and M.A.      ~Barstow                       \inst{\ref{inst:0165}}
\and M.        ~Bellazzini                    \inst{\ref{inst:0043}}
\and V.        ~Belokurov                     \inst{\ref{inst:0001}}
\and N.        ~Blagorodnova                  \inst{\ref{inst:0001}}
\and M.        ~Bonfigli                      \inst{\ref{inst:0222}}
\and A.        ~Bragaglia                     \inst{\ref{inst:0043}}
\and S.        ~Brown                         \inst{\ref{inst:0001}}
\and P.        ~Bunclark$^\dagger$            \inst{\ref{inst:0001}}
\and R.        ~Buonanno                      \inst{\ref{inst:0201}}
\and R.        ~Burgon                        \inst{\ref{inst:0038}}
\and H.        ~Campbell                      \inst{\ref{inst:0001}}
\and R.S.      ~Collins                       \inst{\ref{inst:0091}}
\and N.J.G.    ~Cross                         \inst{\ref{inst:0091}}
\and C.        ~Ducourant                     \inst{\ref{inst:0027}}
\and A.        ~van Elteren                   \inst{\ref{inst:0001}}
\and N.W.      ~Evans                         \inst{\ref{inst:0001}}
\and L.        ~Federici                      \inst{\ref{inst:0043}}
\and J.        ~Fern\'{a}ndez-Hern\'{a}ndez   \inst{\ref{inst:0046}}
\and F.        ~Figueras                      \inst{\ref{inst:0014}}
\and M.        ~Fraser                        \inst{\ref{inst:0001}}
\and D.        ~Fyfe                          \inst{\ref{inst:0165}}
\and M.        ~Gebran                        \inst{\ref{inst:0014},\ref{inst:0553}}
\and A.        ~Heyrovsky                     \inst{\ref{inst:0091}}
\and B.        ~Holl                          \inst{\ref{inst:0012}}
\and A.D.      ~Holland                       \inst{\ref{inst:0038}}
\and G.        ~Iannicola                     \inst{\ref{inst:0201}}
\and M.        ~Irwin                         \inst{\ref{inst:0001}}
\and S.E.      ~Koposov                       \inst{\ref{inst:0001}}
\and A.        ~Krone-Martins                 \inst{\ref{inst:0117}}
\and R.G.      ~Mann                          \inst{\ref{inst:0091}}
\and P.M.      ~Marrese                       \inst{\ref{inst:0201},\ref{inst:0122}}
\and E.        ~Masana                        \inst{\ref{inst:0014}}
\and U.        ~Munari                        \inst{\ref{inst:0002}}
\and P.        ~Ortiz                         \inst{\ref{inst:0165}}
\and A.        ~Ouzounis                      \inst{\ref{inst:0091}}
\and C.        ~Peltzer                       \inst{\ref{inst:0001}}
\and J.        ~Portell                       \inst{\ref{inst:0014}}
\and A.        ~Read                          \inst{\ref{inst:0165}}
\and D.        ~Terrett                       \inst{\ref{inst:0059}}
\and J.        ~Torra                         \inst{\ref{inst:0014}}
\and S.C.      ~Trager                        \inst{\ref{inst:0185}}
\and L.        ~Troisi                        \inst{\ref{inst:0122},\ref{inst:0645}}
\and G.        ~Valentini                     \inst{\ref{inst:0222}}
\and A.        ~Vallenari                     \inst{\ref{inst:0002}}
\and T.        ~Wevers                        \inst{\ref{inst:0032}}
}

\institute{
  Institute of Astronomy, University of Cambridge, Madingley Road, Cambridge CB3 0HA, United Kingdom\relax                                                                                                \label{inst:0001}
\and INAF - Osservatorio Astronomico di Bologna, via Ranzani 1, 40127 Bologna,  Italy\relax                                                                                                                  \label{inst:0043}
\and INAF - Osservatorio Astrofisico di Arcetri, Largo Enrico Fermi 5, I-50125 Firenze, Italy\relax                                                                                                          \label{inst:0024}
\and Institut de Ci\`{e}ncies del Cosmos, Universitat  de  Barcelona  (IEEC-UB), Mart\'{i}  Franqu\`{e}s  1, E-08028 Barcelona, Spain\relax                                                                  \label{inst:0014}
\and Leiden Observatory, Leiden University, Niels Bohrweg 2, 2333 CA Leiden, The Netherlands\relax                                                                                                           \label{inst:0011}
\and STFC, Rutherford Appleton Laboratory, Harwell, Didcot, OX11 0QX, United Kingdom\relax                                                                                                                   \label{inst:0059}
\and Institute for Astronomy, Royal Observatory, University of Edinburgh, Blackford Hill, Edinburgh EH9 3HJ, United Kingdom\relax                                                                            \label{inst:0091}
\and Department of Physics and Astronomy, University of Leicester, University Road, Leicester LE1 7RH, United Kingdom\relax                                                                                  \label{inst:0165}
\and INAF - Osservatorio Astronomico di Roma, Via di Frascati 33, 00078 Monte Porzio Catone (Roma), Italy\relax                                                                                              \label{inst:0201}
\and INAF - Osservatorio Astronomico di Teramo, Via Mentore Maggini, 64100 Teramo, Italy\relax                                                                                                               \label{inst:0222}
\and INAF - Osservatorio astronomico di Padova, Vicolo Osservatorio 5, 35122 Padova, Italy\relax                                                                                                             \label{inst:0002}
\and Department of Astronomy, University of Geneva, Chemin des Maillettes 51, CH-1290 Versoix, Switzerland\relax                                                                                             \label{inst:0012}
\and Laboratoire d'astrophysique de Bordeaux, Universit\'{e} de Bordeaux, CNRS, B18N, all{\'e}e Geoffroy Saint-Hilaire, 33615 Pessac, France\relax                                                           \label{inst:0027}
\and Department of Astrophysics/IMAPP, Radboud University Nijmegen, P.O.Box 9010, 6500 GL Nijmegen, The Netherlands\relax                                                                                    \label{inst:0032}
\and Niels Bohr Institute, University of Copenhagen, Juliane Maries Vej 30, 2100 Copenhagen {\O}, Denmark\relax                                                                                              \label{inst:0035}
\and Centre for Electronic Imaging, Department of Physical Sciences, The Open University, Walton Hall MK7 6AA Milton Keynes, United Kingdom\relax                                                            \label{inst:0038}
\and Serco Gesti\'{o}n de Negocios for ESA/ESAC, Camino bajo del Castillo, s/n, Urbanizacion Villafranca del Castillo, Villanueva de la Ca\~{n}ada, E-28692 Madrid, Spain\relax                              \label{inst:0046}
\and Kavli Institute for Cosmology, University of Cambridge, Madingley Road, Cambride CB3 0HA, United Kingdom\relax                                                                                          \label{inst:0104}
\and CENTRA, Universidade de Lisboa, FCUL, Campo Grande, Edif. C8, 1749-016 Lisboa, Portugal\relax                                                                                                           \label{inst:0117}
\and ASI Science Data Center, via del Politecnico SNC, 00133 Roma, Italy\relax                                                                                                                               \label{inst:0122}
\and Kapteyn Astronomical Institute, University of Groningen, Landleven 12, 9747 AD Groningen, The Netherlands\relax                                                                                         \label{inst:0185}
\and Department of Physics and Astronomy, Notre Dame University, Louaize, PO Box 72, Zouk Mika\"{ e}l, Lebanon\relax                                                                                         \label{inst:0553}
\and Dipartimento di Fisica, Universit\`{a} di Roma Tor Vergata, via della Ricerca Scientifica 1, 00133 Rome, Italy\relax                                                                                    \label{inst:0645}
\and Warsaw University Observatory, Al. Ujazdowskie 4, 00-478 Warszawa, Poland\relax                                                                                                                         \label{inst:0498}
}

\date{Received \textbf{Nov 15, 2016}; accepted \textbf{Dec. 6, 2016}}

 
\abstract
{This paper presents an overview of the photometric data that are part of the first \Gaia\ data release.}
{The principles of the processing and the main characteristics of the \Gaia\ photometric data are presented.}
{The calibration strategy is outlined briefly and the main properties of the resulting photometry are presented.}
{Relations with other broadband photometric systems are provided. The overall precision for the \Gaia\ photometry is shown to be at the milli-magnitude level and has a clear potential to improve further in future releases.}
{}


\maketitle


\section{Introduction}
The ESA \Gaia\ satellite mission, launched in December 2013, started its full-sky astrometric, photometric and spectroscopic survey of the Milky Way Galaxy in July 2014 \citep{GaiaMission}. The first large-scale publication of data from the mission took place on 14 September 2016, when positions and broadband magnitudes were published for 1.14 billion stars. This paper gives an overview of the photometric data that are part of that data release.

The \Gaia\ photometric data consists of three systems: broadband \gmag\ (350 - 1000~nm) magnitudes, optimised for the astrometric data to collect a maximum of light, but also provide accurate integrated flux measurements over the mission; integrated fluxes in a blue \BP (330 - 680~nm)  and red \RP (640 - 1000~nm) channel, obtained from the integration of dispersion spectra; and low-resolution BP and RP dispersion spectra. The wavelengths coverage is still based on the pre-mission specification \citep{2010A&A...523A..48J}, a full calibration of the actual passbands will be carried out as part of the next \Gaia\ data release. The photometric data presented here and included in \Gaia\ Data Release 1 (DR1) only concerns the G band and was obtained from the processing of data collected during the first 14 months of the \Gaia\ mission. A more comprehensive description of the data included in \GDR1 is presented in \cite{GaiaDR1}. It should be noted that this is only a first reduction, with a partial external calibration, and a number of issues in the data and the calibration models still to be resolved. 

This paper provides a general overview of the photometric processing. It is accompanied by three other papers describing specific aspects of the photometric processing in much more detail: a paper on the calibration principles \citep{PhotPrinciples}, one on the technical issues presented by the processing of the \Gaia\ photometric data and on the solutions implemented 
\citep{PhotProcessing} and one on the extensive validation activities that preceded the release \citep{PhotValidation}.
Among the papers accompanying \GDR1, the two papers \cite{VarProc} and \cite{CephRRL} show the huge potential and exquisite quality of the \Gaia\ photometry.  

Section~\ref{sec:inputdata} describes the input data with an emphasis on the use and implications for the data presented here. The calibration models and principles of the calibrations are presented in Sect.~\ref{sec:calmod}. Precision and internal consistency of the data is described in Sect.~\ref{sec:precacc}. In Sect.~\ref{sect:photsyst} we show some comparisons between the \Gaia\ G band and other photometric systems. A brief summary of results is given in Sect.~\ref{sect:summ}.

\section{The input data \label{sec:inputdata}}

This section provides a brief overview of the input data from which the \Gaia\ photometry has been derived. \Gaia\ observations are obtained through CCDs operating in time-delayed integration (TDI) mode, integrating the signal as it moves across a CCD in approximately 4.5 seconds. A single pixel has an integration time of $\approx 0.001$~s, corresponding to a movement along scan of $\approx 0.06$ arcsec. Integration times can be shortened for bright stars by using so-called gates. The shortest gate, as used for the brightest stars, integrates over just 4 pixels. Pixels in the across-scan direction are about 0.18 arcsec. The movement across scan for a full CCD transit can vary between $\pm 4$ across-scan pixels. We refer to \cite{GaiaMission} and \cite{PhotPrinciples} for a more detailed description of the instrument and the data. 

\subsection{The sky mappers \label{ssec:sm}}


\Gaia\ scans the sky, more or less along great circles, with two telescopes with overlapping fields of view on the focal plane. This is true for all CCDs except for the two Sky Mapper CCDs \citep[for a view of
the focal plane layout, see][]{PhotPrinciples}, which provide a preliminary detection of a source and assign a provisional brightness. Sources are first detected by Sky Mapper 1 (SM1) for the preceding field of view and Sky Mapper 2 (SM2) for the following field of view. The SM strips contain seven CCDs each. 

All \Gaia\ CCDs can be operated in a gated mode, where only a section of the CCD is integrated, thus effectively reducing the exposure time. This is designed to reduce the occurrence of saturated images, ultimately allowing for stars as bright as magnitude 3 to be observed. In the SM1 and SM2 CCD strips charges are integrated over about half the CCD width, this corresponds to gate 12 being permanently active with a resulting effective exposure time of 2.9 seconds.
The data are accumulated onboard in samples of 2 by 2 pixels. With the full width half maximum of the point spread function at about 1.8 pixels (0.1 arcsec), it is clear that the images on the SM detectors are undersampled. There is no further compensation for bright images, which as a result tend to be saturated. For the faintest stars (down to magnitude 20.7) the resolution of the data as sent back to Earth is further reduced to 4 by 4 pixels. The different binning strategies are referred to as window classes. The SM data are always provided in the form of 2D images. This is important for analysing data on close binary stars, although that has not yet been implemented at this stage of the data reductions. A full overview of the different window classes and gates can be found in Table~1 in \cite{PhotPrinciples}.

Twenty-eight individual calibrations (two strips, seven rows and two window classes) are required to fully characterise the SM data. The photometric processing of these data uses the background-corrected fluxes as derived in the image parameter determination in the initial data treatment \citep[fitting observed counts to a calibrated point-spread function, see also][]{IdtRef}.  

In conclusion, the SM data provide (low angular resolution) 2D image information, without the disturbances caused by the overlapping fields of view and without the problems of linking data obtained with different gate settings (affecting the other instruments), but suffer from lower resolution and saturation for the brighter images. Given the significantly higher noise level on the SM data, these data have not been included in the accumulated G band photometry in \GDR1.

\subsection{The astrometric field \label{ssec:af}}

The astrometric field consists of 9 strips of 7 CCDs, referred to as AF1 to AF9. In total there are 62 AF CCDs, as one CCD in the AF9 strip is used as wave-front sensor. The AF1 strip is special, as it has the onboard task to confirm or reject the detections made by the SM strips. The  provisional brightness  assigned to the source by the SM is implemented in the AF1 detection through gate and window settings. The transit confirmation by AF1 forms input to the onboard attitude control system, which determines, amongst others, the scan rates in the two fields of view. This defines the recordings by the other CCD 
strips. Gate setting and data sampling have been adapted for the AF1 strip, for example the samples in the 2D windows are 1$\times$2 pixels, instead of 1$\times$1 pixels as in the other AF strips, which often leads to numerical saturation (caused by the A/D converter). The pixel saturation instead is largely avoided by adjusting the gate settings to the across-scan saturation levels of each CCD. Within a single field-of-view transit a bright star very frequently encounters different gate settings (thus different effective integration times) per CCD.

In total, 8 different gate settings are used for the AF CCDs. In addition, data are transmitted as either 2D images for stars brighter than magnitude 13, or, for fainter stars, compressed to 1D images in AF1, and read out as 1D images in AF2 to AF9. Only sources bright enough to be observed as 2D images can trigger the activation of a gate. The 1D images, in which the pixels in the across-scan direction are added up, can be either a strip of 12 or 18 single samples, the number depending on the brightness of the source. These settings are referred to as window classes. Each setting of gate and window class creates a different instrument, with its own calibration. Each of the 62 CCDs thus create 10 instruments per field of view, giving a total of 1240 individual calibrations units. In fact, every pixel column (i.e. the array of pixels aligned to the scan direction) in the CCD could be considered a different instrument because of different sensitivity: this is partly taken care of by the small-scale calibration (Sect.~\ref{ssec:sscal}), while the large-scale calibration is responsible for variations in the mean response CCD-to-CCD and for each field of view (Sect.~\ref{ssec:lscal}). \cite{PhotPrinciples} provides more details about how this is carried out.

In conclusion, the AF1 data often shows numerical-readout saturation rather than pixel saturation and a lower across-scan resolution and image quality for 2D images. The remaining 55 CCDs are comparable in the data produced. However, towards strip AF9 a small fraction of images are lost when the across-scan motion in the field of view moves them out of the area of the CCD. 

\subsection{The photometers, integrated fluxes \label{sec:photflux}}

The blue and red photometers (BP and RP) concern two strips of 7 CCDs each with prisms intercepting the light 
beam to produce low-resolution dispersion spectra. The pre-processing of these data is part of the photometric processing chain, and concerns bias and background corrections. The integrated flux, after correcting for bias and background, is the input to the integrated \BP and \RP flux calibrations.

Next to the total integrated flux, integrated fluxes over fixed wavelength ranges of the spectra are obtained. These are referred to as the spectral shape coefficients (SSCs) and play an important role in the calibration models \citep[see][]{PhotPrinciples}. They represent pseudo-filters, which allow for more detailed spectral dependencies to be modelled. The computation of these require the calibration of effects due to geometry and dispersion function to allow the conversion of sample positions into absolute wavelengths. 

The photometers use 5 different gate settings and two different window classes, in total 6 calibrations per CCD and field of view, giving a total of 84 calibrations per strip and 168 calibrations in total. In addition to this, there are 4 SSCs for each of the two passbands, each of which requires a further 84 
calibrations, giving a total of 336 calibrations. 

\section{Calibration models and strategy \label{sec:calmod}}

The photometric calibrations described here are internal calibrations, in other words, they have been designed to define a photometric system based on the internal consistency of the sky. This is later followed by an external calibration which, by means of a relatively small number of specially selected spectral calibration stars, tries to reconstruct the actual photometric passbands of the \Gaia\ photometric system.

The  internal photometric calibrations consist of two main elements: a large-scale calibration to represent variations in the telescope and detectors on timescales of about a day, and a small-scale calibration to represent local variations in CCD response. 

\subsection{Colour dependencies \label{ssec:colourdep}}

The colour dependencies in both the large- and small-scale calibration models represent in first instance the differences in quantum efficiency between the CCDs and optics. During the early stages of the mission, however, it became apparent that mirror-contamination development was a major factor in the detector chain, causing wavelength-dependent flux loss and therefore requiring colour-dependent flux corrections that evolve significantly with time \citep{PhotPrinciples}. 

Colour coefficients derived from the broadband integrated \BP and \RP fluxes are insufficient to reflect and model the changes in response for different spectral types, in particular to distinguish between the effects of temperature and surface gravity. For this purpose we have introduced the SSCs derived from the BP and RP dispersion spectra. Even though the SSCs are in no way clean spectral passbands, owing to the image smearing effects, they do provide a more detailed representation of the way spectra of different types are affected by passband variations.

\subsection{Large-scale flux calibrations \label{ssec:lscal}}

Each calibration referred to above is known as a calibration unit. In total there are 2080 of these units per time interval, which is typically one day.  In \GDR1, while SM fluxes are also calibrated, they do not contribute to the mean photometry, so only observations from  1744 calibration units are used. All these calibrations only concern large-scale effects. Large-scale effects represent the influence of variations in the telescope, such as focus drift and mirror condensation. They also represent part of the CCD-specific large-scale response variations. For a five-year mission this amounts to of order 3 million calibrations, the monitoring and quality control of which is a major effort on its own.

Accidental calibration units occur in case of unexpected gates (when a window is affected by a gate triggered by a simultaneously observed bright source), complex gates (when only part of a window is affected by a gate triggered by other sources) or truncated windows (when two windows overlap and the window with lower priority, typically the faintest one, is truncated). These cases do not happen very often and the signal retrieved also tends to be rather weak. There are therefore not enough observations to enable a calibration model to be determined and applied. In \GDR1 these cases are not yet being treated, with serious consequences in crowded regions, where nearly one-third of images are lost because of truncated windows. 

\subsection{Small-scale calibrations \label{ssec:sscal}}

The small-scale calibrations represent the local response variations on the CCDs, generally related to the manufacturing process. Because of this there is no distinction between the two fields of view, but all other aspects of calibration units remain the same. The small-scale calibrations try to model in detail the response variations of the CCDs as a function of the across-scan coordinate. Known small-scale discontinuities are related to so-called stitch blocks, which are small rectangular areas on the CCD. 

As the small-scale variations are linked to the CCDs rather than the telescope, it is possible to use data accumulated over long time intervals (many months to a year). For the current \GDR1 publication, all 14 months of data was used in a single set of small-scale calibrations.

\subsection{The calibration strategy \label{ssec:calstrat}}

For photometry, as with astrometry, \Gaia\ is considered to be self-calibrating. The sky, or at least a large percentage of the objects observed, are non-variable down to a little above the accuracy levels achieved by \Gaia, in other words, a few mmag at most. After calibration the observed fluxes of such assumed not-variable sources should be the same within the standard uncertainty margins, independent of the time of observation or the applicable calibration unit. The internal calibration therefore aims at creating an internal photometric system consistent over the entire instrument using all suitable sources. This system is achieved by solving the large- and small-scale calibrations in an iterative process, as these two kinds of calibrations cannot be obtained simultaneously, requiring data sets with very different time lengths. The process is described in detail in \cite{PhotPrinciples}. This system can then be connected to an external, for example ground-based, system using a small number of external calibration stars especially observed for this purpose. These stars are referred to as spectro-photometric standard stars \citep[SPSSs, see also][]{Spss}.

The internally calibrated system has to cover all proper gate and window combinations, all CCDs and both fields of view. To link all these different calibration units, there needs to be overlap, i.e.\ a significant fraction of sources has to have been observed in different calibration units. Some overlap is readily obtained, from the transit of a star over 9 CCDs across the focal plane. The CCDs often use different gate settings (but not different window settings) to compensate for local saturation effects. Other connections, such as those between different field of views and between different rows of CCDs, are provided by the scanning law, by observing the same stars at different scan angles. Some linking between 1D and 2D windows is supported by so-called calibration faint stars, which are relatively faint stars that would normally have been assigned 1D windows and are occasionally (in up to 0.4\% of cases) observed in a non-gated 2D window. Another main contributor to the overlapping calibration units is the inaccuracy of the onboard magnitudes, in particular when the images have reached saturation on the SM detectors. In those cases onboard magnitude estimates can reach uncertainty levels of 0.5 magnitudes. This means that many stars with magnitudes in the range 7 to 12 are observed with a range of gate settings, and provide the overlap that is needed to build the internally consistent photometric system.  Similarly sources with magnitude close to 13 and 16 may be observed with different window classes, although at these magnitudes the onboard magnitude estimate is much more accurate and therefore the mixing is not as efficient as in other cases thus complicating the calibration process \citep[see][]{PhotPrinciples}.

\section{Precisions and accuracies \label{sec:precacc}}

Precisions of the photometric data are defined by
\begin{enumerate}
\item the photon statistics of the observed image and its read-out noise;
\item the accuracy of the flux recovery from the observed image;
\item the accuracy of the calibration model for the recovered fluxes.
\end{enumerate}

The accuracy (Cramer-Rao) limit is set by the photon statistics of the observed counts, which includes background contributions, and excludes any counts that are affected by saturation. There are in addition various sources of readout noise that need to be taken into account. Where the background level is relatively low, and no saturation is seen, the estimated errors on the parameters derived from the image typically are proportional to the square root of the total photon count. This applies to photometric and astrometric parameters derived from the image. Background contributions for the faintest stars and saturated pixels for the brightest stars increase these estimated errors.

The flux recovery from the image depends on the accuracy of the predicted image shape, which is the applicable effective point spread function (PSF) for 2D images, or line-spread function (LSF) for 1D images. The effects of the colour of the source need to be taken into account, as well as variations of the PSF 
with positions in the field of view. However, this is still not the actual PSF for an image projected onto the focal plane. This is also affected by sampling in the along and across scan directions, focal-plane drift across the scan direction, and local CCD response variations. Some, but not all, of these effects can be incorporated in the predicted PSF. What remains causes small systematic errors in the flux estimates that should be resolved in the photometric reductions. At this early stage of the mission, however, the PSF profiles are still under development and no dependencies other than CCD and field of view are taken into account. This leaves some significant systematics that will have to be dealt with for future releases. One other consequence of the limited accuracies of the PSF profiles is the poor goodness-of-fit statistics for the image fit. As a result, the observed standard deviation of the image fit contains a significant calibration-error contribution, which is not Gaussian and cannot easily be corrected for.

The final stage is the actual photometric reduction. Here we face two main challenges, the changing conditions of the telescope during the first year of observations and the detailed variations in the CCD response as a function of the across-scan coordinate. For some gate settings there is a third challenge: 
the small number of observations per unit of time. Shortcomings in these calibrations leave a small calibration error, than mainly affect the brightest stars. The final error contribution comes from the linking of the different calibration units, which depends on the accuracies of the individual calibrations and can thus be affected by the calibration-error contributions. These calibrations are described in detail in \cite{PhotProcessing}. The validation of the results is described in \cite{PhotValidation}.

Once the internal system is fully settled, and internal estimated errors are well understood, the link to an external system can be made \citep{PhotPrinciples}. This involves a reconstruction of the representative passband of the internal system. With the relatively strong evolution of the telescope during the first year of observations, this definition of the representative passband is not unambiguous. Therefore, for \GDR1, the determination of the passband was postponed and the nominal pre-launch instrument response was used instead, so that the magnitude zero point and a set of transformations to other external system could be provided. More details can be found in the \href{http://gaia.esac.esa.int/documentation/GDR1/Data_processing/chap_cu5phot/sec_phot_calibr.html}{online documentation for \GDR1}.

\section{The \Gaia\ photometric system\label{sect:photsyst}}

\begin{figure}[t]
\centering
\includegraphics[width=8.5cm]{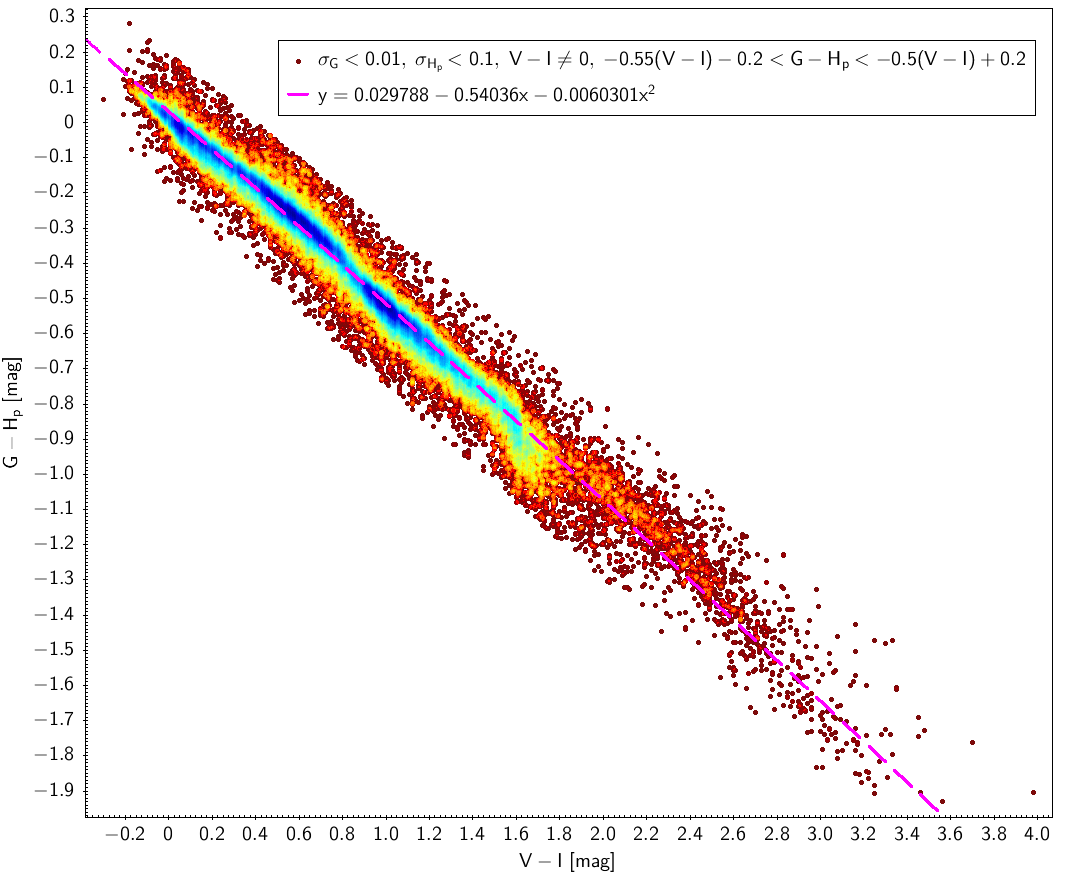}
\caption{Comparison between the \Gaia\ and \Hip\ broadband magnitudes, as a function of the colour index $V-I$. The colours indicate the density of stars, from red (low) to blue (high). Outliers, most of which can be traced back to double stars, were removed.}
\label{fig:ghpvi}
\end{figure}

The broadband \gmag\ magnitude in \Gaia\ is derived from the calibrated flux \citep[see also][]{PhotValidation} as
 \begin{equation}
G = -2.5\log({\rm flux}) + {\rm zp}
\end{equation}
where ${\rm zp}$ is the photometric zero point derived by the external calibration. For a detailed discussion on the determination and values of the zero point see \cite{PhotPrinciples}.

Preliminary transformations have been determined between the \Gaia\ and other commonly used photometric systems. A range of preliminary empirical transformations was determined, and compared with results expected from the theoretical passband definitions. Small deviations were found, but the overall behaviour is close to the expectations. 

The \Hip\ $H_p$ magnitudes \citep{Hipparcos} are most similar to the \gmag\ broadband. The comparison between the two systems is shown in Fig.~\ref{fig:ghpvi}. The relation between the magnitudes in the two systems is given by
\begin{equation}
G - H_p = 0.0029788	- 0.54036 (V-I) - 0.0060301 (V-I)^2
\end{equation}
The range of applicability of this relation is $-0.2<V-I<3.5$.
The standard deviation over this range is 0.040 mag, which, based on the accuracies of the \Hip\ and \Gaia\ photometry, implies that there are significant secondary dependencies present in this relation.

More transformations between the \Gaia\ photometric system and other systems were calculated and are available in the \href{http://gaia.esac.esa.int/documentation/GDR1/Data_processing/chap_cu5phot/sec_phot_calibr.html}
{online documentation for \GDR1}. The coefficients of the transformations to the most common photometric systems are also listed in Table~\ref{table:coeffs}.
\begin{table*}[t]
\begin{center}
\caption{Coefficients of the colour-colour polynomial transformations between \Gaia\ and \tyctwo, Johnson-Cousins and SDSS photometric systems. An indication of the range of applicability of these
relations is given in the 7th column. Additional details on the source selection criteria for the definition of the empirical relations are available in the online documentation.}
\label{table:coeffs}
\begin{tabular}{lrrrrrcc}
\hline\hline
& & $(B_T-V_T)$& $(B_T-V_T)^2$&$(B_T-V_T)^3$ &$\sigma$& Applicability range & Tycho-2\\
$G-V_T$&  0.0079363	& -0.4235 & 0.10048	& -0.070742 & 0.066 & $-0.2<B_T-V_T<2.0$ &  \citep{Tycho2}\\
\hline
& & $(V-I)$& $(V-I)^2$&$(V-I)^3$ &$\sigma$& & Johnson-Cousins\\
$G-V$&  0.02266	& -0.27125 & -0.11207 & & 0.028	& $-0.25<V-I<3.25$ &  \citep{2012PASP..124..140B}\\
\hline
  & & $(g-i)$& $(g-i)^2$&$(g-i)^3$ &$\sigma$& & SDSS\\
$G-g$& -0.098958 & -0.6758 & -0.043274 & 0.0039908 & 0.028 & $-0.4<g-i<3.0$ &  \citep{SDSSDR10}\\
\hline
\end{tabular}
\end{center}
\end{table*}

\section{Summary of results\label{sect:summ}}

We present a brief summary of the characteristics of the photometric data as presented in the first \Gaia\ data release. A more detailed description can be found in \cite{PhotValidation}.

\subsection{Distribution over the sky}

\begin{figure*}[ht!]
\centering
\includegraphics[width=130mm]{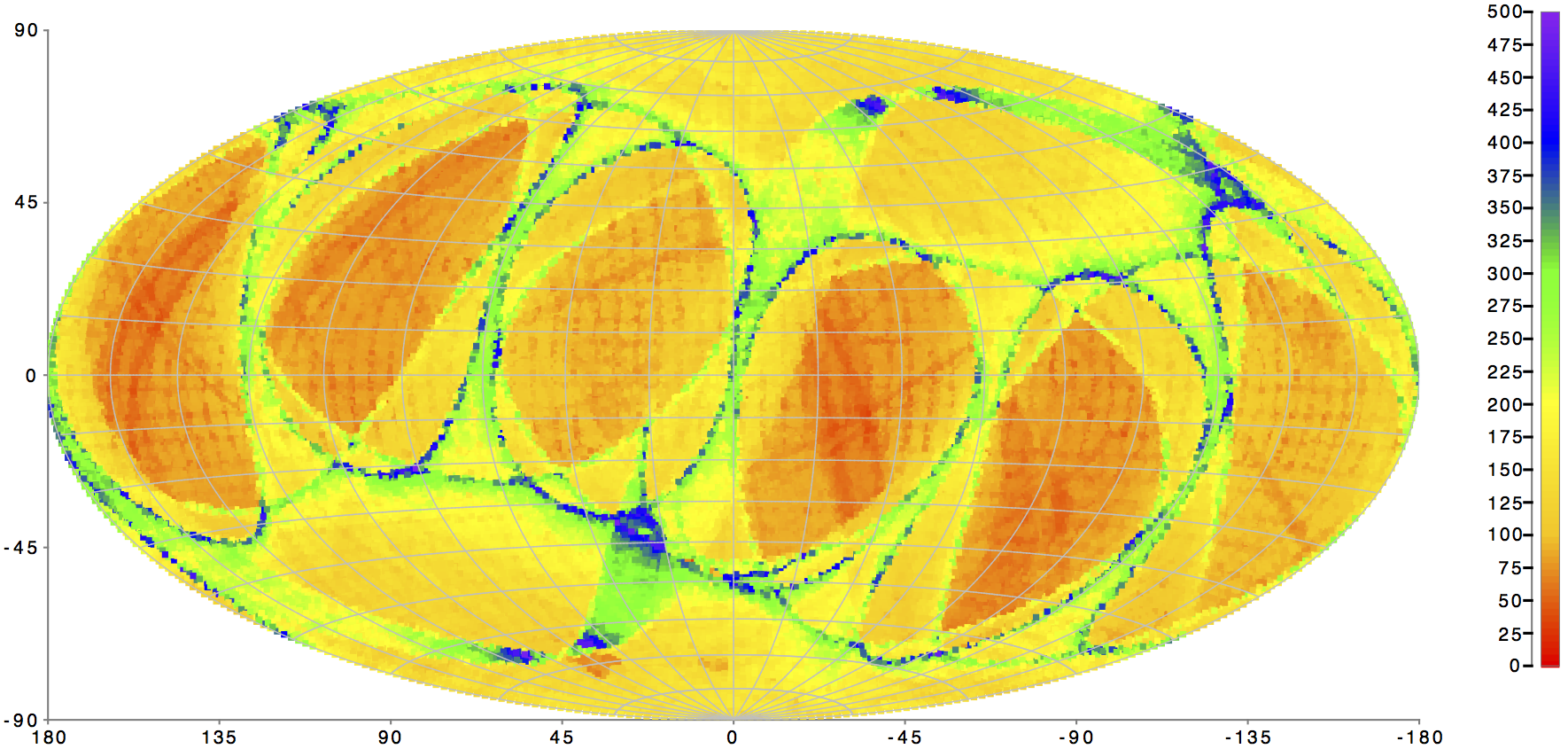}
\caption{Sky distribution of the average number of CCD transits per HEALPix pixel (at level 5, about 3.36 square degrees per pixel) for window-class 1 observations (1D, long windows, stars with estimated brightness between 13 and 16 magnitudes). Approximately 8 to 9 CCD transits correspond to one field-of-view transit. Positions in the sky are given in equatorial coordinates using a Hammer-Aitoff equal-area projection. The large number of transits near the ecliptic poles are due to the special scanning mode used during the first 4 weeks of the mission.}
\label{fig:nobssky}
\end{figure*}

\begin{figure*}[ht!]
\centering
\includegraphics[width=130mm]{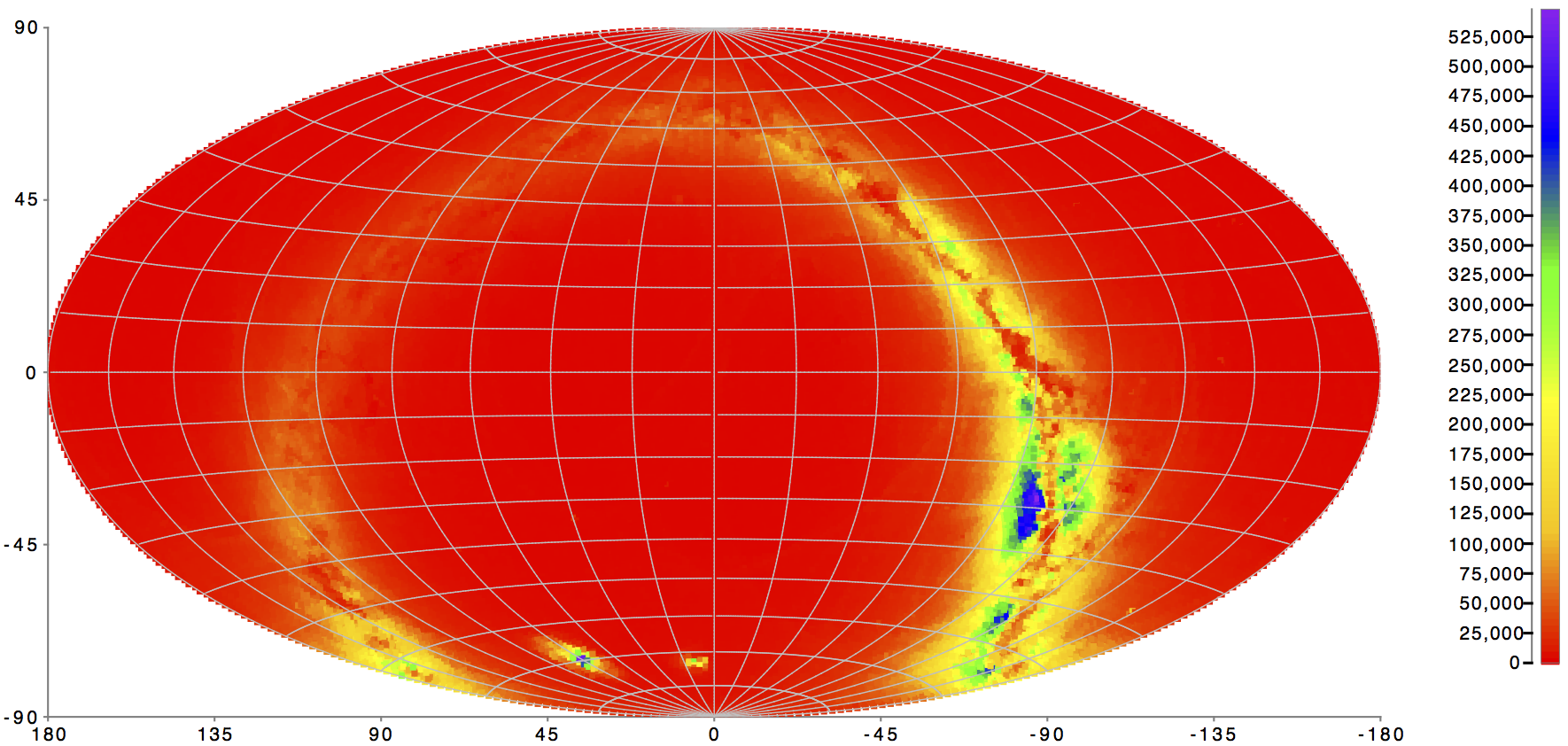}
\caption{Sky distribution of the number of sources per HEALPix level 5 pixels \citep{healpix}. Positions in the sky are given in equatorial coordinates using a Hammer-Aitoff equal-area projection.}
\label{fig:densitysky}
\end{figure*}

The distribution of the number of observations and number of sources on the sky, as shown in Fig.~\ref{fig:nobssky} and \ref{fig:densitysky}, expose some of the extremes of the conditions under which the photometric calibrations 
have to operate. The variation of the number of observations depends on the way \Gaia\ scans the sky. The scan is necessarily built around the ecliptic plane, keeping the spin axis of the satellite at a fixed angle of 45\degr from the direction of the Sun \citep[see also][]{GaiaMission}. The precession of the spin axis around the direction of the Sun and the rotation of the satellite provide a complete sky coverage in at least two scan directions every 6 months. Gaps in the coverage can be seen as darker red bands, and are caused by various interruptions of the data stream, some of which will be recovered at a later stage in the reductions \citep[see][]{GaiaDR1}.

The distribution of sources, with the extremes in density near the Galactic plane, potentially creates other complications for the reductions. The distribution over colour indices can be very different depending on 
the area of the sky under examination, with both extremely reddened faint stars and very young bright blue stars in the Galactic plane. At this early stage, with some of the calibration models still imperfect, variations in the colour distributions and the density of sources can lead to unwanted variations of the 
colour coefficients. These variations do not reflect the evolution of the instrument. This increases the calibration noise in the mean photometric data. Once the colour dependencies are better understood, and the calibration model represents the instrument more accurately, these errors will rapidly decrease. 

\begin{figure}[t]
\centering
\includegraphics[width=8cm]{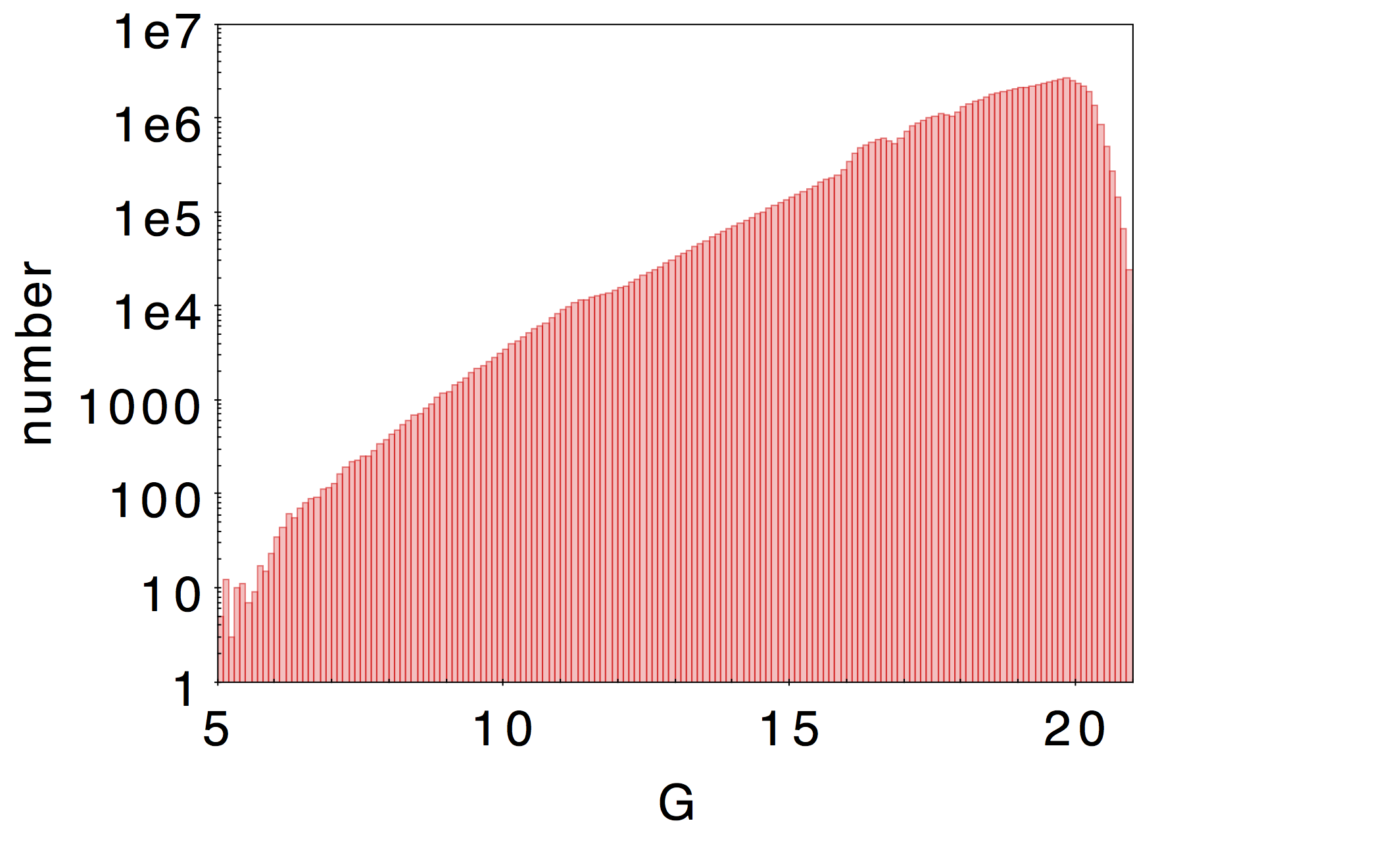}
\caption{Number of sources per magnitude (bin width 0.1 mag), with about 100 CCD transits recorded. Features seen towards the fainter magnitudes are thought to be the result of data download priorities. The magnitude scale is the calibrated G band magnitude.}
\label{fig:accumul100}
\end{figure} 

The completeness as a function of magnitude of the photometric data in this first \Gaia\ data release is affected by a range of external factors, such as the priorities defined for downloading data from the satellite, and difficulties in the initial data treatment, particularly related to poor performances in periods of high data volume. In both cases data for faint stars was selectively lost \citep[see for more details][]{DPACP-16}. When the problem was caused by the on-ground processing, the data can, and will, be recovered in the next processing cycles. A total of about 1.14 billion sources were processed successfully. Around 0.1 per~cent of the data (about 1.2 million stars) is not included in this release becasuse it is either too red or to blue to fit within the calibration boundaries. This, again, is a temporary 
feature, caused by the requirements in the initialisation of the reference fluxes \citep{PhotProcessing}. Figure~\ref{fig:accumul100} shows the distribution over magnitude of sources with about 100 observations ($69~025~678$ sources with number of CCD observations between 95 and 105). Some peculiar features in the magnitude distribution in the range $16<G<20$ are the result of the download priorities and upstream processing issues mentioned earlier.
 
\subsection{Overview of precision and internal consistency}

\begin{figure}[t]
\centering
\includegraphics[width=7.5cm]{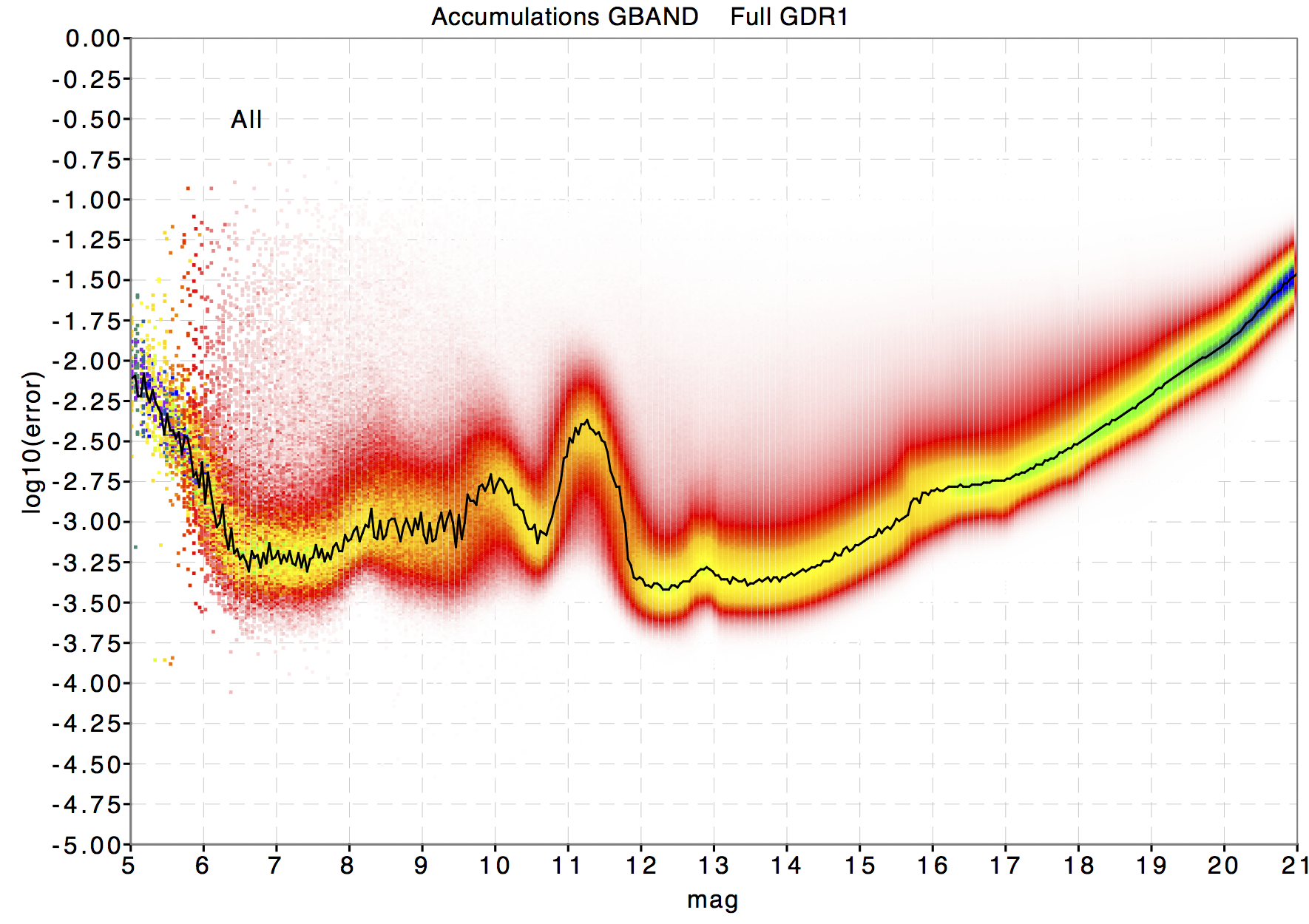}
\caption{Chart showing the estimated uncertainties on the weighted mean $G$-band values as a function of magnitude. The colours represent the density of sources, from red (low) to blue (high). The distribution of sources has been flattened for this plot to make sure that features at all magnitudes were visible. For stars between magnitudes 12 and 16 the relation between estimated error and magnitude is close to what could be expected based on noise models. For fainter stars the effect of the background can be noted, while for brighter stars the estimated errors are larger than expected due to effects related to different gate
settings and saturation. For a comparison with the expected errors see \cite{PhotValidation}.}
\label{fig:gmagsigm}
\end{figure}

This section presents an overview of what is essentially the precision or internal consistency of the photometric data. Figure \ref{fig:gmagsigm} shows the distribution of the estimated standard uncertainties on the mean \gmag\ magnitude as a function of the mean G. The estimated standard uncertainties on the mean were determined from the standard deviation of the calibrated observations and the number of observations used. When that number is small, the standard deviation may be underestimated. Some effects are easily recognisable, for instance the effect of photon statistics, which is the main source of error for the faintest stars. The effect of saturation is evident for sources brighter than 
$\sim6.5$, while the three "bumps" in the middle are caused by the gates. Finally, some residual effects of linking inaccuracies can be seen at $G\sim13$ and $G\sim16$. The reason for this is that stars with magnitudes close to a 
gate or window transition have measurements obtained on either side of that transition due to inaccuracies in onboard magnitude estimates, thus artificially increasing the estimated error on the mean magnitude if the calibrations 
have not been entirely successful in converging to a consistent system covering all instrument configurations.

\begin{figure}[t]
\centering
\includegraphics[width=7.5cm]{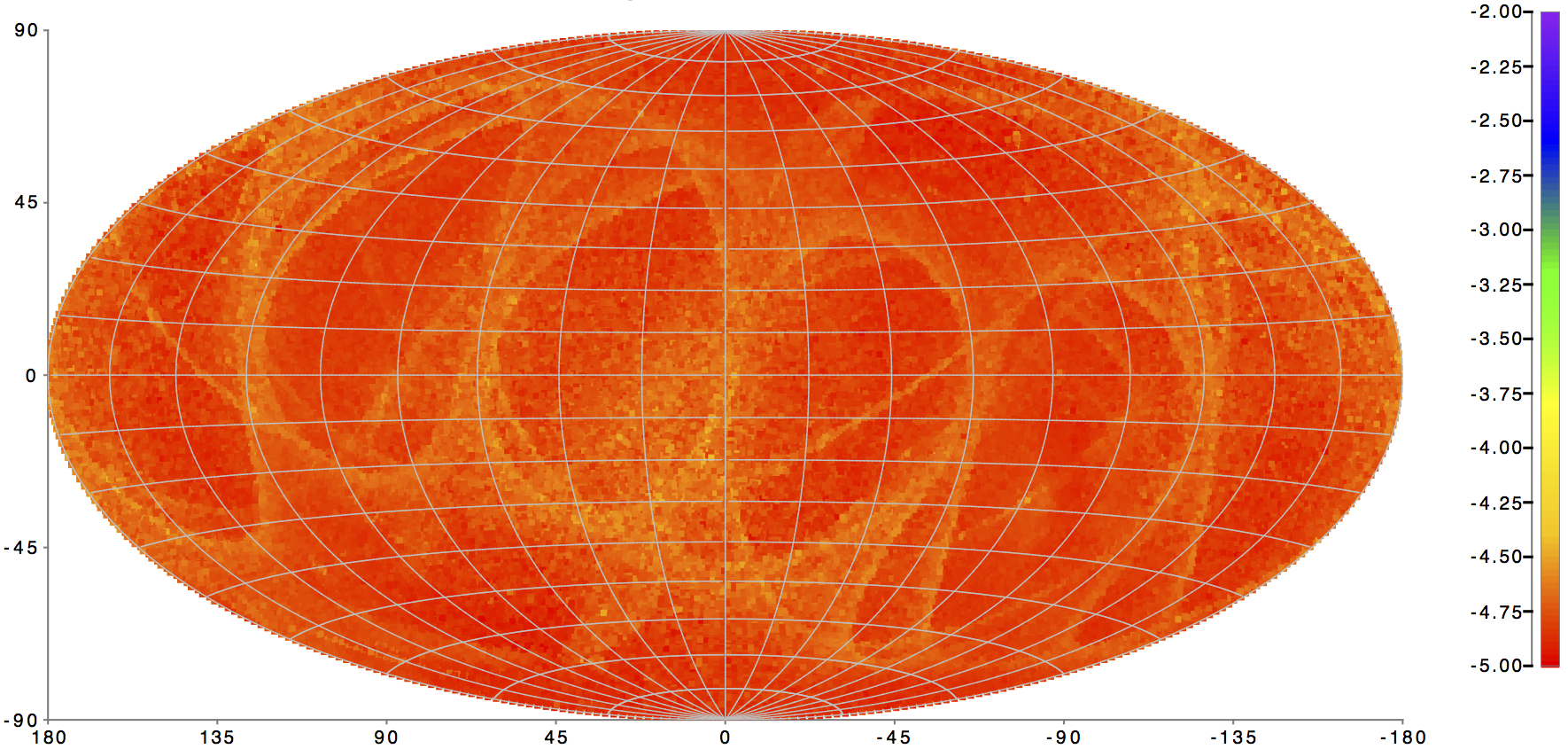}
\includegraphics[width=7.5cm]{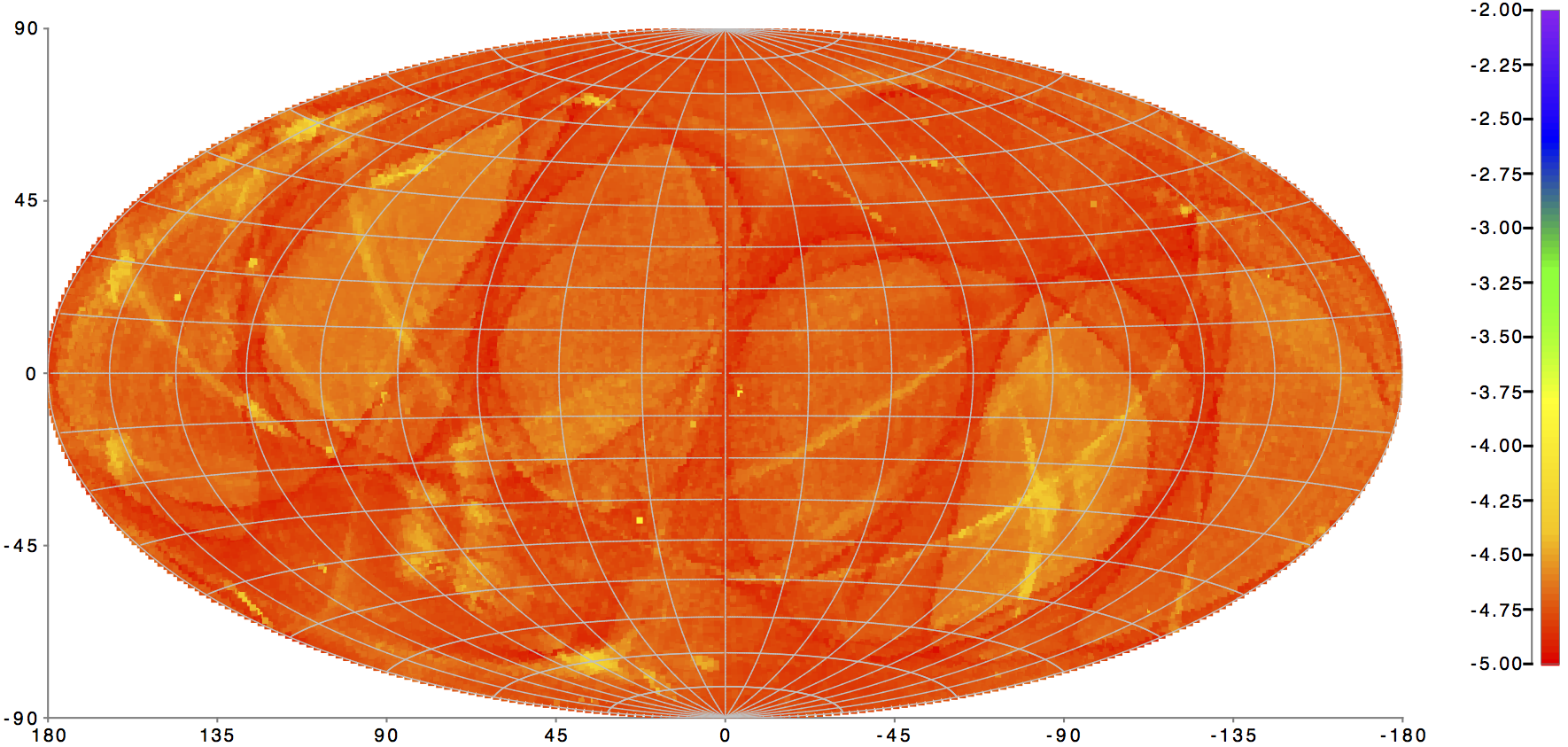}
\includegraphics[width=7.5cm]{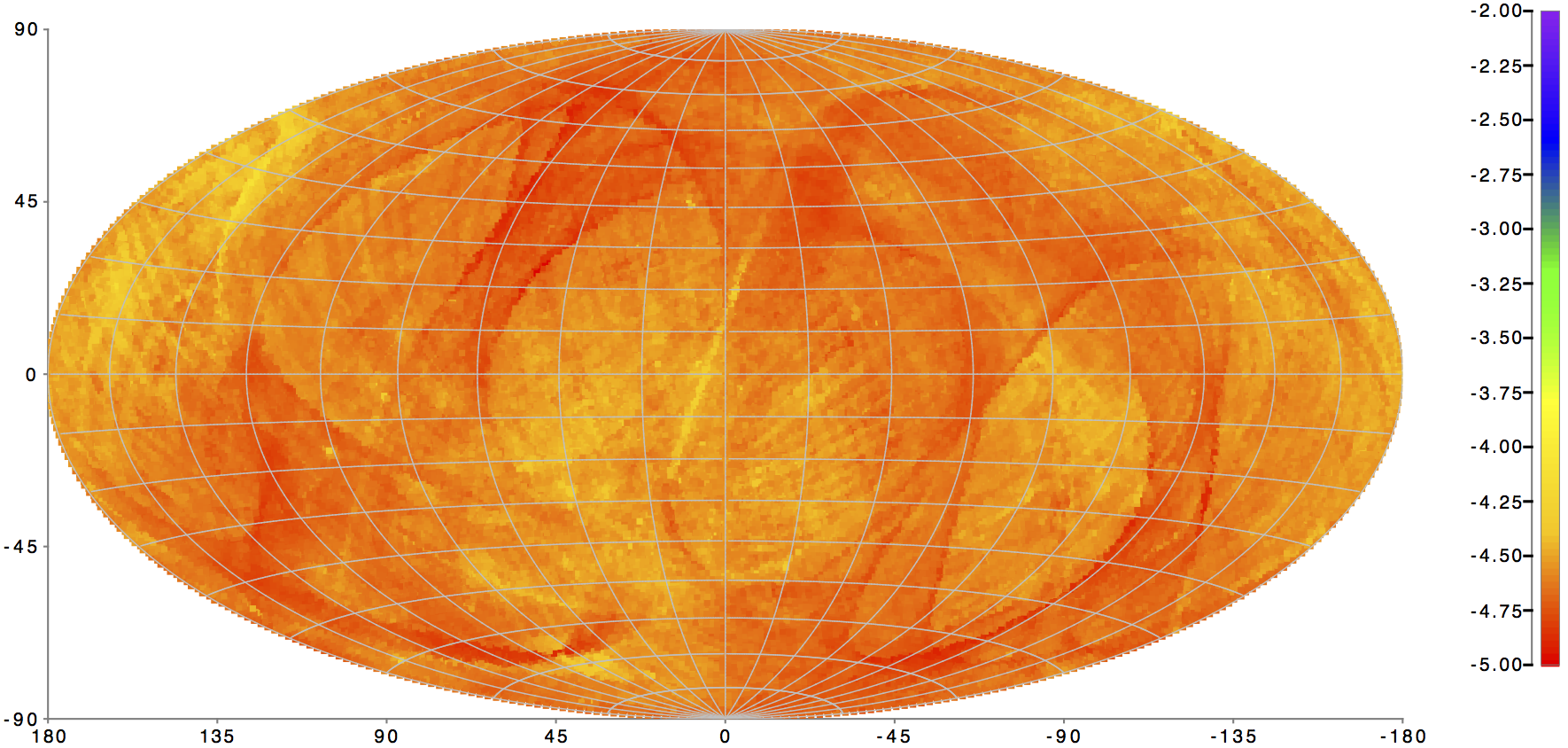}
\caption{Sky distribution of the estimated errors on the mean \gmag\ photometry for the different window classes. From top to bottom: Window Class 0 (2D windows), Window Class 1 (1D windows), Window Class 2 (smaller 1D windows). The window class is allocated depending on the magnitude of the observed star as determined onboard during detection.}
\label{fig:errormaps}
\end{figure}

The plots in Fig. \ref{fig:errormaps} show the sky distribution of the estimated errors on the mean \gmag\ photometry for different window classes. All three plots show distributions clearly correlated with the scanning law caustics. However, the distributions look somehow different between Window Class 0 (bright stars, 2D windows, top plot) and the other two classes (fainter stars, 1D windows, 18 or 12 samples long, central and bottom plots). For the fainter sources the estimated errors are smaller when more observations are available, although some great circle regions show poor estimated errors (possibly caused by a single problematic calibration period). The pattern is somehow inverted for bright sources, showing inferior estimated errors when more observations are available. This is not yet fully understood and could be the result of low numbers of observations when mean magnitudes are used in the calibrations. In a self-calibrating system this may locally result in fitting the errors on the observations when very few observations are available. Nevertheless the accuracy even in these cases reaches the mmag level.
 
\section{Conclusions}

This paper presents the photometric data included in the first \Gaia\ data release. Only Gband photometry is included in this release. 

A high-level summary of the photometric data and processing is given here. The accompanying papers provide details as follows: \cite{PhotPrinciples} for the detailed definition of the calibration models,  \cite{PhotProcessing} for a description of the software solutions and processing strategies, and \cite{PhotValidation} for an in-depth analysis of the results of the photometric processing. 

Not all instrumental effects are calibrated at this early stage in the mission and the photometry published in \GDR1 is the result of the first cyclic processing with no iteration between the various systems in the \Gaia\ Data Processing and Analysis Consortium.

Some calibration effects have not yet been fully understood and there seem to be some systematics at the 10 mmag level particularly around $G=11$ \citep[see][]{PhotValidation}.
The \Gaia\ photometric catalogue covers the entire sky, providing measurements of the average brightness of sources down to magnitude 21 (at different levels of completeness at this stage) in a single photometric system. The overall accuracy of the photometric data reaches the $3-4$ mmag level.

Future releases will enhance this photometric catalogue with improved G band photometry and the addition of colours and low-resolution spectra for all sources thus creating the most complete and accurate photometric catalogue to date. 

\begin{acknowledgements}
This work was supported in part by the MINECO (Spanish Ministry of Economy) - FEDER through grant 
ESP2013-48318-C2-1-R and MDM-2014-0369 of ICCUB (Unidad de Excelencia 'Mar\'ia de Maeztu').

We also thank the Agenzia Spaziale Italiana (ASI) through grants ARS/96/77, ARS/98/92, ARS/99/81, 
I/R/32/00, I/R/117/01, COFIS-OF06-01, ASI I/016/07/0, ASI I/037/08/0, ASI I/058/10/0, ASI 2014-025-R.0, 
ASI 2014-025-R.1.2015, and the Istituto Nazionale di AstroFisica (INAF).

This work has been supported by the UK Space Agency, the UK Science and Technology Facilities Council.

The research leading to these results has received funding from the 
European Community's Seventh Framework Programme (FP7-SPACE-2013-1) 
under grant agreement no. 606740.

The work was supported by the Netherlands Research School for Astronomy (NOVA) and the Netherlands Organisation for Scientific Research (NWO) through grant NWO-M-614.061.414.

The work has been supported by the Danish Ministry of Science.
\end{acknowledgements}

\bibliographystyle{aa} 
\bibliography{refs} 

\end{document}